\documentclass[final,1p,times,authoryear]{elsarticle}
\pdfoutput=1

\usepackage{amssymb}
\usepackage{lipsum}
\usepackage{slashed}

\journal{J Phys A}

\newcommand{\BE}{\begin{equation} \begin{array}{c}}
\newcommand{\EE}{\end{array}\end{equation}}

\newcommand{\BT}{\begin{theorem}}
\newcommand{\ET}{\end{theorem}}

\newcommand{\gcpl} {\texttt{g}}

\newcommand{\LAG}{\mathcal{L}}

\newcommand{\SG}{{\sigma}}
\newcommand{\SB}{{\overline{\sigma}}}
\newcommand{\psiB}{{\overline{\psi}}}

\newcommand{\ZB}  { \overline{Z}}
\newcommand{\dd} {\texttt{d}}
\newcommand{\DD} {\texttt{D}}
\newcommand{\kk} {\texttt{k}}
\newcommand{\pp} {\texttt{p}}

\usepackage{xcolor}
\newcommand{\blue}[1]{#1}

\begin{document}

\begin{frontmatter}



\date{\today }

\title {
   Conformal invariance of antisymmetric tensor field theories\\ in any even dimension.
}


\author[a]{Jean Thierry-Mieg}
\ead{mieg@ncbi.nlm.nih.gov.}
\author[b,c,d]{Peter Jarvis}
\ead{peter.jarvis@utas.edu.au}

\affiliation[a]{
  National Library of Medicine, National Institute of Health, 
  8600 Rockville Pike, Bethesda MD20894, U.S.A.
}
\affiliation[b]{
  School of Natural Sciences (Mathematics and Physics),
  University of Tasmania, Private Bag 37,
  Hobart, Tasmania 7001, Australia.
}
\affiliation[c]{
  Alexander von Humboldt Fellow.
}
\affiliation[d]{
  Corresponding author.
}

\begin{abstract}
    Using a theorem of Jackiw and Pi expressing the delicate balance of the spin and the orbital momentum,
    we systematically classify the flat-space massless Lagrangian quantum field
    theories that are invariant under the global conformal group $SO(\DD,2)$.
  We recover in a uniform way the facts that scalars and spinors are invariant in any dimension, and that gauge $\pp$-tensors
  are invariant only in $2\pp+2$ dimensions. This case includes the Maxwell theory in 4 dimensions and the Kalb-Ramond 2-forms
  theory in 6 dimensions.
  We then construct two new classes of Lagrangians extending the Avdeev-Chizhov self-dual tensor model to higher dimensions,
  one class using a symmetric metric and the other a skew metric in internal space. Finally,
  we prove in the same uniform way that both classes are
  conformal invariant in any even dimension. In 4 dimensions, these self-dual tensors naturally couple to the
  chiral Fermions of the standard model.
\end{abstract}




\end{frontmatter}




\section{Introduction}
\label{sec:Intro}

The aim of this article is to study the conformal invariance of Lagrangians describing
antisymmetric tensor fields in any even dimension
using the elegant \blue{criterion} formulated by  \cite{Jackiw_2011}, \blue{see also \cite{CallanColemanJackiw1970}}.

Despite the
fact that the conformal group plays a crucial role in the superstring models
which cover all dimensions from 2 to 10, there is, to our knowledge, no systematic
general classification of the flat space conformally invariant Lagrangian field theories. To fill this gap
we provide here a direct and uniform proof for the known cases
of the scalars, spinors vectors, and the Kalb-Ramond gauge tensors (KR) theories
and then extend our proof
 to two new generalizations in $\DD$ dimensions
of the \cite{AvdeevChizhov1994A,AvdeevChizhov1994B} self-dual tensors.
The first class (AC) uses a symmetric metric in internal space and the second class (CP), an
anti-symmetric metric.
We call it (CP) because such Lagrangians flip sign under (C) charge or (P) parity flips, 
but are invariant under Landau (CP) symmetry. We prove that the massless
scalars and spinors are conformal invariant in any dimension, that the  (KR)
Lagrangians describing antisymmetric tensors with $\pp$ indices
are conformal invariant only in dimension $\DD = 2 \pp + 2$
and then the new result that the (AC) and (CP) Lagrangians describing self-dual tensors with $\pp$ indices
are conformal invariant in any even dimension $\DD = 2 \pp$.
Notice that \cite{Witten2004} and followers study an entirely different kind of theories
where rather than the tensor $T$ itself, its curvature $dT$ is self-dual.
Such models are not part of our classification because they do not admit a Lagrangian.
On the other hand, these authors
do not consider our new self dual classes (AC) or (CP).

To understand the literature dealing with conformal invariance, 
three symmetry groups must be carefully distinguished: (A) the 1 dimensional global scale transformations,
 (B) the $(\DD+2)(\DD+1)/2$ dimensional global conformal group $SO(\DD,2)$ which combines
the dilatations,
the $\DD(\DD-1)/2$ Lorentz rotations, the $D$ Poincar\'e
translations, and the $D$ boosts, in total 15 parameters in $\DD=4$ dimensions, and
(C) the infinite dimensional Weyl local scale transformations which apply to curved space.

Although they are undoubtedly different, these three groups are strongly related.
By definition, (B) and (C) contain (A). Furthermore the three groups
have interconnected applications in physics.
For example, one objective of the seminal physics reports by \cite{Nakayama2015} was to establish
that global scale invariance (A), combined with unitarity, implies
conformal invariance (B), which, according to \cite{Farnsworth2017}, in turn implies Weyl invariance (C).
And in the sixties, as recalled in the Nakayama's introduction,
Migdal could not easily convince the great Bogolioubov that conformal
invariance (B) was more powerful than scale invariance (A).

However, full equivalence is not established: for example,
\cite{Karananas2016} show that not all scalar conformal invariant theories (B)
can be made locally scale invariant (C).
Several other studies, for example \cite{Showk2011,Wu2017,Nakayama2017,Lee2021},
show that there exist (A) scale invariant and  unitary theories which are not (B) conformally invariant.

A contrario, the interplay between these three groups is sometimes overlooked.
For example, in a paper titled '... the Cheshire cat smile ...', \cite{Shapiro2024}
has studied an antisymmetric tensor model in curved space
invariant under local rescaling (group C: the cat)
\blue{and shown it to be one-loop renormalizable}.
In the flat space limit,
although local scale invariance disappears, \blue{the model remains renormalizable}.
Shapiro compares 
this phenomenon to the Cheshire cat, missing
that the tail of the cat remains visible. Indeed, in that limit,
his model coincides with the Avdeev-Chizhov model (AC). Therefore,
it is not
only scale invariant (group A: the smile of the cat), but
\blue{as shown below in section \ref{sec:ATF},} it is also
conformally invariant (group B: the tail of the cat).
Although we have no proof, it is not inconceivable that, as in this model of Shapiro,
the flat space limit of any locally scale invariant field theory (C)
should be globally conformal invariant (B).

Let us now sketch the organization of our analysis. We only consider Minkowski flat space.
In dimension $\DD$ any Lorentz invariant free field Lagrangian $\LAG(\Phi,\partial \Phi)$
is automatically scale invariant. We just need to choose the canonical dimension $\dd$ of each field
such that the dimension of each propagator has total dimension $D$, counting as 1 the dimension of the
derivatives. In a Bosonic model with two derivatives, like the scalar or the Maxwell theory, $\DD = 2 \dd + 2$.
In a Fermionic Dirac Lagrangian with one derivative $\DD = 2 \dd + 1$. In a Bosonic quartic model (appendix B),
$\DD = 2 \dd + 4$. Couplings can then be added as long as the coupling constants are dimensionless.

Na\"{i}vely, it may then seem that any globally (A) scale invariant field theory should be (B) conformally invariant, but this is not true.
As formalized by \cite{Jackiw_2011}, the problem lies in the influence
of the conformal transformations
on field derivatives because the field derivatives are \blue{local but} not tensorial, meaning that
their transformation rules depend
on the derivatives of the coordinate transformations.
This affects the apparent orbital momentum and spin of the particles. But for the theory to be invariant,
these two modifications must exactly compensate each other, otherwise the apparent rotational energy of the fields
would not be conserved. This constrains the structure of the field propagators and of all couplings involving
field derivatives, as they must exquisitely match
the internal spin of the particles.

In section \ref{sec:Conformal_section}, we recall the theorem of \cite{Jackiw_2011} and systematically prove that the classic scalar
and spin-half free Lagrangians are conformally invariant in any dimension, that the Maxwell Lagrangian is invariant only in $\DD=4$
dimensions, (\cite{Bateman1910}), and that covariant derivatives and Yang-Mills-Higgs couplings are conformally
invariant in $\DD=4$ because the coupling constants are dimensionless. We also discuss the spin 3/2 supergravity field.

In section \ref{sec:ATF}, we prove that the free scale invariant (A)  and gauge invariant massless
Kalb-Ramond (KR) $\pp$-form Lagrangians $\LAG = ({}^*dT)\wedge (dT)$ are conformally invariant (B)
only in dimension $\DD = 2 \pp$ (see below for the notations). 
Although we have here provided the benefit of a systematic proof, the results thus far are known.
We then turn to new results.
We study the \cite{AvdeevChizhov1994A,AvdeevChizhov1994B} Lagrangian (AC),
which describes self-dual $2$-forms in $4$ dimensions,
generalize it to $\DD$ dimensions, and
further generalize it to a new anti-symmetric (CP) model, first encountered in our recent study of
the $SU(2/1)$ superalgebraic structure of the electroweak interactions, (\cite{Thierry_Mieg_Jarvis_2021}).
We prove that the (AC) and the (CP) models are both conformally invariant (B) in any even dimension.

In the short section \ref{sec:DF}, we discuss the physical content of antisymmetric tensor field theories
  and count the degrees of freedom in $\DD$ dimensions.
We also show that in $\DD=4$ dimensions, the triple scalar-vector-tensor coupling $\Phi F^{\mu\nu}T_{\mu\nu}$
which involves derivatives ($F_{\mu\nu} = \partial_{[\mu}A_{\nu]}$) is conformal invariant.

In the appendices, we analyze 2 additional toy models discussed in the literature.
\blue{In \ref{ap:N6},} we show in a simpler way that the gauge
fixed Maxwell model of \cite{Showk2011}
is conformally invariant. \blue{In \ref{ap:MQ}, we show}
that contrary to section C.3 of \cite{Lee2021},
the scale invariant quartic Maxwell model
$F^{\mu\nu} \Box F_{\mu\nu}$
is not conformally invariant. 

\section{Conformal invariance}
\label{sec:Conformal_section}

A basis of the generators of the $SO(D,2)$ conformal Lie algebra in $\DD$ dimensions with Minkowski metric
$g_{\mu\nu}$
is provided by the $SO(D-1,1)$ Lorentz transformations
$M_{\mu\nu}$,
the dilatation $\Delta$, the translations $P_{\mu}$ and the conformal boosts $K_{\mu}$
with commutation relations (see for example \cite{Showk2011}, equation A.1).
\BE
\label{eq:2.0}
[M_{\mu\nu},M_{\rho\sigma}]= i g_{\nu\rho}M_{\mu\sigma} - i g_{\nu\sigma}M_{\mu\rho} - i g_{\mu\rho}M_{\nu\sigma} + i g_{\mu\sigma}M_{\nu\rho}\;,\;
  \\\;
[M_{\mu\nu},P_{\rho}] = i g_{\nu\rho}P_{\mu} - i g_{\mu\rho}P_{\nu}\;,\;\;  [M_{\mu\nu},K_{\rho}] = i g_{\nu\rho}K_{\mu} - i g_{\mu\rho}K_{\nu}\;,\;\;
    \\\;
[P_{\mu},K_{\nu}] = + 2i (g_{\mu\nu}\Delta - M_{\mu\nu})\;,
\\\;
    [\Delta,M_{\mu\nu}] = 0\;,\;\;   [\Delta,P_{\mu}] = -i P_{\mu}\;,\;\;   [\Delta,K_{\mu}] = - i K_{\mu}
    \;.
    \EE
\noindent
  The algebra acts on fields $\Phi$ with canonical scaling dimension $\dd$ as
\BE
\delta \Phi = (\epsilon^{\mu\nu}(-i(x_{\mu}\partial_{\nu} - x_{\nu}\partial_{\mu}) + \Sigma_{\mu\nu})
-i\; \epsilon^{\mu}\partial_{\mu}
\\
-i\;  \epsilon'^{\mu}(2x_{\mu}\dd +2i x^{\alpha} \Sigma_{\alpha\mu} + 2x_{\mu}x^{\alpha}\partial_{\alpha}-X^2\partial_{\mu})
-i\; \epsilon (x^{\mu}\partial_{\mu} + \dd)
) \Phi\; 
\EE
where the $\epsilon^{\mu\nu}$ parameterize the Lorentz transformations $M_{\mu\nu}$\, the operator $\Sigma_{\mu\nu}$ acts on the spin of
the field $\Phi$  \blue{(see the definitions in equation (\ref{eq:2.3}) below)}, the $\epsilon^{\mu}$ parameterize the Poincar\'e translations $P_{\mu}$,
the $\epsilon'^{\mu}$ the conformal boosts $K_{\mu}$\, and finally $\epsilon$ parametrizes the dilatation $\Delta$.
The caveat is that the action of the boosts on the fields is non trivial.

\cite{Jackiw_2011}, elaborating on the earlier work of \cite{CallanColemanJackiw1970},
 have proposed a very elegant criterion to 
 assess the conformal invariance of a globally scale invariant and Poincar\'{e} invariant Lagrangian.
 The necessary and sufficient condition is that the field virial
\BE
\label{eq:2.1}

V_{\sigma} = \displaystyle{\frac {\partial \LAG}{\partial (\partial_{\alpha} \Phi)}} \;(\dd\;g_{\alpha\sigma} - \Sigma_{\alpha\sigma})\;\Phi
\EE
summed over all fields $\Phi$ should be a divergence,
\BE
\label{eq:2.2}
V_{\sigma} = \partial^{\alpha} \omega_{\sigma\alpha}\;.
\EE
In this definition, $\LAG$ is the Lagrangian, $g_{\mu\nu}$ is the metric, $\dd$ is the canonical dimension of the fields,
and $\omega$ is an arbitrary local polynomial in the fields and their derivatives.

To use this criterion, we need to compute $\dd$ and to normalize $\Sigma$ properly.
On a scalar field, $\DD = 2 \dd + 2$ and $\Sigma$ vanishes.
On a chiral spinor, $\DD = 2 \dd + 1$ and the rotations are represented by the Dirac matrices.
On a tensor, $\DD = 2 \dd + 2$ and each vector index rotates like the coordinates.
\BE
\label{eq:2.3}
\Sigma_{\alpha\beta}(\phi) = 0\;,\;\;\;   \Sigma_{\alpha\beta}(A_{\mu}) = - g_{\beta\mu}A_{\alpha} + g_{\alpha\mu}A_{\beta},
\\
\Sigma_{\alpha\beta}(\psi) = \frac{1}{4}(\gamma_{\alpha}\gamma_{\beta} - \gamma_{\beta}\gamma_{\alpha})\;\psi\;,
\\
\Sigma_{\alpha\beta}(T_{\mu\nu ...}) = (-g_{\beta\mu}T_{\alpha\nu ...} + g_{\alpha\mu}T_{\beta\nu ...}) +
\\
+(g_{\beta\nu}T_{\mu\alpha ...} - g_{\alpha\nu}T_{\mu\beta ...}) + ... 
\EE
\blue{where $(...)$ denotes the iteration over all the indices of the tensor $T$.}
The rotation of a fully antisymmetric $\pp$-tensor $T$, i.e. a $\pp$-form, involves $\pp$ pairs of terms.
As a first test, we compute the field virial $V_{\sigma}$ for a real free massless scalar field
\BE
\label{eq:2.4}
\LAG = -\frac{1}{2} \partial_\mu\Phi\;\partial^{\mu}\Phi\;,\\
V_{\sigma}(\Phi) = \displaystyle{\frac {\partial \LAG}{\partial(\partial_{\alpha} \Phi)}} \;(\dd\;g_{\alpha\sigma})\;\Phi =-\frac{1}{2}\;\dd\;\partial_{\sigma} \Phi^2\;.
\EE
We found a divergence, therefore the scalar Lagrangian is conformally invariant in any dimension. We now try the 
massless spinors
\BE
\label{eq:2.5}
\LAG =  i \psiB \gamma^{\mu}\partial_\mu\psi\;,\;\;\;
\\
V_{\sigma}(\psi) = \displaystyle{\frac {\partial \LAG}{\partial(\partial_{\alpha} \psi_L)}} \;(\dd\;g_{\alpha\sigma}\;-\Sigma_{\alpha\sigma})\;\psi
\\
= i \psiB \gamma_{\sigma} (\dd - (\DD-1)/2)\psi = 0\;.
\EE
Here, the canonical dimension $\dd$ where $\dd=(\DD-1)/2$ is exactly compensated
by the contraction of the Dirac matrices $\gamma^{\alpha}(-\gamma_{\alpha}\gamma_{\sigma}+\gamma_{\sigma}\gamma_{\alpha})/4= \gamma_{\sigma}\;(-\DD + (2-\DD))/4= \gamma_{\sigma}\;(-\DD+1)/2$.
So the field virial of the spinors vanishes, hence the Dirac Lagrangian is conformal invariant
in any dimension.
By the same argument, the Rarita-Schwinger spin $3/2$ Lagrangian $\psiB_{\mu} \slashed{\partial} \psi_{\mu}$,
where we used the constraint
$\gamma^{\mu}\psi_{\mu}=0$ to eliminate the spin $1/2$ components, is also conformally invariant in any dimension, because the constraint eliminates the contribution of the vector index rotation.
We now
examine the Maxwell Lagrangian 
\BE
\label{eq:2.6}
\LAG = - \frac{1}{4} (\partial_{\mu}A_{\nu} - \partial_{\nu}A_{\mu})(\partial^{\mu}A^{\nu} - \partial^{\nu}A^{\mu})\;,\;\;\;
\\
\\
V_{\sigma}(A) = \displaystyle{\frac {\partial \LAG}{\partial(\partial_{\alpha} A_{\mu})}} \;(\dd\;g_{\alpha\sigma}\;-\Sigma_{\alpha\sigma})\;(A_{\mu})
\\
=
- (\dd - 1) A^{\alpha} (\partial_{\sigma} A_{\alpha}  - \partial_{\alpha} A_{\sigma})\;. 
\EE
The field virial vanishes only in dimension $(\dd=1 \Leftrightarrow \DD=4)$,
hence the Maxwell theory is conformal invariant only in 4 dimensions,
\cite{Bateman1910,Jackiw_2011,Showk2011}.
Actually, it is also secretly invariant in $\DD=3$ because then
the theory is degenerate and $F_{\mu\nu}$
is dual to a scalar (same references).
In $\DD = 4$ dimensions, we could turn in Yang-Mills interactions
or include in the Lagrangian terms in $\Phi^4$, $A^4$, or $A^2\Phi^2$,
this would introduce covariant derivatives on the right hand sides
without altering the conformal invariance since, in the definition of the virial (\ref{eq:2.1}),
we only differentiate relative to
the derivatives of the fields.
Yet mass terms are forbidden because they break the global scale invariance of the theory.
In $\DD > 4$ dimensions, the coupling constant $\gcpl$ in the covariant derivative
$D_{\mu} = \partial_{\mu} + \gcpl A_{\mu}$ has negative dimensions, also breaking conformal invariance.

\section {Antisymmetric tensor fields}
\label{sec:ATF}
Antisymmetric tensor fields $T_{\mu\nu...}$ of rank $\pp$ are equivalent to $\pp$-forms
\BE
\label{eq:1.1}
T = (1/p!) \;T_{\mu\nu ...}dx^{\mu}\wedge dx^{\nu}...
\EE
where $\wedge$ denotes the antisymmetric exterior product of the $1$-forms $dx^{\mu}$,
\blue{and $(...)$ the iteration over all indices of the tensor $T$.}
They were first discussed in quantum field theory in relation with models
for higher spin fields, see for example
\cite{Ogievetsky1967notoph}, and they play an important role
in supergravity and superstring theory. The most natural Lagrangian ($KR$),
often associated in physics to the names of \cite{KalbRamond1974}
\BE
\label{eq:1.2}
\LAG^{KR}\;d^{\DD}x = \frac{1}{2} ({}^{*}dT)\wedge (dT)
\EE
is a straightforward generalization of the Maxwell Lagrangian $\LAG^{M}\;d^4x = {\scriptstyle 1/2} \;{}^*dA\wedge dA$,
where $A = A_{\mu}dx^{\mu}$ denotes the Maxwell 1-form,
$d$ the Cartan exterior differential, ${}^*$ the Hodge dual, and $d^{\DD}x$ the volume element $\DD$-form.
Since $d$ is nilpotent, $d^2=0$, the Lagrangian $\LAG^{KR}$ is gauge invariant
under a transformation of the $\pp$ form $T$ parameterized by a $(\pp-1)$-form $\xi$:
\BE
\label{eq:1.3}
\delta T^{(\pp)} = d\xi^{(\pp-1)}\;\;\Rightarrow\;\;\delta \LAG^{KR} = 0\;.
\EE

However, this is not the only possibility. In a stunning development, Avdeev and Chizhov ($AC$) 
discovered in 1994 a new Lagrangian describing in 4 dimensions 
a conformally invariant massless self-dual 2-tensor naturally interacting
with chiral Fermions
\blue{since in 4 dimensions the
minimal coupling $\overline{(\psi_R)}_L \;\SG^{\mu}\SB^{\nu}T_{\mu\nu}\;\psi_L$
is automatically self dual : ${}^*(\SG^{[\mu}\SB^{\nu]}) = -i \,\SG^{[\mu}\SB^{\nu]}$, }
\cite{AvdeevChizhov1994A,AvdeevChizhov1994B}. The phenomenology is well developed in \cite{Chizhov2011}.

Here, we study these 2 types of tensors in any dimension.
We consider the Kalb-Ramond gauge tensor fields ($KR$) in dimension $\DD$,
we generalize the known $\DD=4$ Avdeev-Chizhov matter tensor fields ($AC$) to higher dimensions, and we study the conformal invariance
of these theories. In addition, we construct a new class of Lagrangians using a skew metric in the internal space that we call ($CP$) for two reasons.
We prove that our new Lagrangian $\LAG^{CP}$ is $C$: conformally invariant and
we show that it switches sign under the separate action of $C$: charge conjugation and $P$: parity conjugation,
but remains invariant under the combined Landau $CP$: charge-parity conjugation.

The $\LAG^{CP}$ Lagrangian was first encountered as the tensor propagator 
in our recent study of the $SU(2/1)$ superalgebraic structure of the electroweak interactions (\cite{Thierry_Mieg_Jarvis_2021}).
By checking in the next section the scalar-vector-tensor triple coupling present in that model,
we finally show that the complete $SU(2/1)$ super-chiral super-connection Lagrangian is conformally invariant
in 4 dimensions.

\noindent

Let us now consider the two standard tensor Lagrangians, Kalb-Ramond and Avdeev-Chizhov, as provided at the very beginning of
the second paper on this subject by \cite{AvdeevChizhov1994B}, rediscovered in curved space by \cite{Shapiro2024} (his equations 11 and 13)
\BE
\label{eq:2.7}
\LAG^{KR} = \frac{1}{4} [(\partial_{\alpha}T_{\mu\nu})(\partial^{\alpha}T^{\mu\nu}) - 2 (\partial^{\alpha}T_{\alpha\mu}) (\partial_{\beta}T^{\beta\mu}) ]
\;, 
\\
\LAG^{AC} = \frac{1}{4} [(\partial_{\alpha}T_{\mu\nu})(\partial^{\alpha}T^{\mu\nu}) - 4 (\partial^{\alpha}T_{\alpha\mu}) (\partial_{\beta}T^{\beta\mu}) ]
\;.
\EE

Up to a total derivative, the trivial Lagrangian $T_{\mu\nu} \Box T^{\mu\nu}$ is among their linear combinations.
But it is not conformally invariant because it treats
the 6 components of $T_{\mu\nu}$ as six independent scalars without recognizing their tensorial spin structure. Therefore
the Kalb-Ramond and the Avdeev-Chizhov Lagrangians cannot both be conformally invariant.
Generalizing these equations, let  $T$ be a $\pp$-form and let us compute the field virial
associated to the d'Alembertian ($T \Box T)$  and to the square of the
divergence $(\partial . T)^2$ in $\DD$ dimension
\BE
\label{eq:2.8}
\LAG^{\Box} = -\frac{1}{2}\; (T_{[\mu\nu ...]})\;\Box (T^{[\mu\nu ...]})
\;,\\
V_{\sigma}^{\Box} = \dd (\partial_{\sigma} T_{[\mu\nu ...]}) \; T^{[\mu\nu ...]} - 2\pp \;(\partial_{\alpha}T^{[\alpha\nu ...]}) \;T_{[\sigma\nu...]}
\;,
\\
\\
\LAG^{\nabla} = \frac{1}{2}\; (\partial^{\alpha}T_{[\alpha\nu ...]})\;(\partial_{\beta}T^{[\beta\nu...]})
\;,\\
V_{\sigma}^{\nabla} = (\DD -\dd - \pp) \;(\partial_{\alpha}T^{[\alpha\nu ...]}) \;T_{[\sigma\nu...]}
\;.
\EE
The first term in $V_{\sigma}^{\Box}$ is a divergence and can be dropped, but the other 2 terms are not.
  However they can be combined. We can choose $k$ such that
the linear combinations $V^{k}_{\sigma} = V_{\sigma}^{\Box} - \kk\; V_{\sigma}^{\nabla}$
of these two field virials becomes a divergence:
\BE
\label{eq:2.9}
V^{\kk}_{\sigma} = V_{\sigma}^{\Box} - \kk\; V_{\sigma}^{\nabla} = \partial_{\sigma}(...)\;.
\EE
The necessary and sufficient condition is that that $\kk$ solves
\BE
\label{eq:2.10}
2\pp - \kk (\DD - \dd - \pp) =  0
\;.
\EE
Therefore in this case and only in this case the corresponding Lagrangian $\LAG^{\kk} = \LAG^{\Box} - \kk \LAG^{\nabla}$
is conformally invariant.
To apply this formalism, we work modulo integration by parts, denoted $\cong$,  and rewrite the Maxwell-Kalb-Ramond Lagrangian
in the desired form 
\BE
\label{eq:2.11}
\LAG^{KR}\;d^{\DD}x = \frac{1}{2} {}^*(dT) \wedge (dT) \cong \frac{1}{ \pp !} (\LAG^{\Box} - \pp \LAG^{\nabla})\;d^{\DD}x\;.
\EE
So $\LAG^{KR} \cong \LAG^{\kk}$ if we choose the linear combination $\kk=\pp$
and the $KR$ theory is conformally invariant  if and only if $2+\dd-\DD+\pp=0,$ implying $\pp = \DD - \dd -2 = \dd$. In other words, the
canonical dimension $\dd$ of the tensor $T$ must be equal to its rank $\pp$. The solutions are a zero-form in $\DD=2$, i.e. 
the scalars of string theory, a $1$-form in $\DD=4$, i.e. Maxwell electromagnetism, a $2$-form in $\DD=6$, and so on.
But the $\DD=4$ Kalb-Ramond gauge 2-form Lagrangian is not conformally invariant. In 6 dimensions we could have a
2-form cubic coupling $(dT)^2 + T^3$ and maintain conformal invariance.
These results coincide with the very different proof of \cite{Witten2004}.

Let us now analyze in more detail the Avdeev-Chizhov model.
We know that in dimension $\DD=2p$, the antisymmetric $\pp$-tensor $T_{[\mu\nu ...]}$ does not form an irreducible representation
of the Lorentz group.
Denoting ${}^*T$ the Hodge dual of the tensor $T$,
  we can split $T$ it into its Minkowski self-dual $Z$ and anti-self-dual part $\ZB$ as follows
\BE
\label{eq:2.12}
Z = T + i{}^*T\;,\;\;
\ZB=  T - i{}^*T,\;\;{}^*Z = - i Z\;,
\\
T = (Z + \ZB)/2\;,\;\; {}^*T = -i(Z - \ZB)/2\;.
\EE
These equations are valid in Minkowski space where ${}^{**}T = -T$. In Euclidean space, drop the $i$ and adjust the signs.
Now, consider the chiral Lagrangian
\BE
\label{eq:2.13}
\LAG^{\chi} = \frac{1}{2(p-1)!}  (\partial .  \ZB)(\partial .  Z) = \frac{1}{2(\pp-1)!} (\partial^{\alpha}\ZB_{\alpha\mu ...})(\partial_{\beta}Z^{\beta\mu ...})\;.
\EE
\blue{where $(...)$ denotes the iteration over all the indices of the self-dual tensor $Z$.}
Using integration by parts, we can again rewrite this Lagrangian in the desired form
\BE
\label{eq:2.14}
\LAG^{\chi} \cong \frac{1}{ \pp !} ( \LAG^\Box  - \DD \LAG^{\nabla})
\EE
which coincides in $\DD=4$ dimensions with the Avdeev-Chizhov Lagrangian (\ref{eq:2.7}) 
and provides a generalization in any even dimension. So $\kk = \DD = 2\pp$ and $\dd =(\DD-2)/2=\pp-1$,
hence condition (\ref{eq:2.10}) $2\pp - \kk (\DD - \dd - \pp) = 0$ holds
and therefore the generalized Avdeev-Chizhov theory is conformally invariant in any even dimension.
We recover $\pp=1$ in $\DD=2$, i.e. the pseudo-scalars of string theory, $\pp=2$ in $\DD=4$, i.e. the original Avdeev-Chizhov Lagrangian, and so on.

This simple structure of the chiral Lagrangian $\LAG^{\chi}$ provides a way to construct the antisymmetric pseudo-scalar
Lagrangian whose existence was proven in \cite{Jarvis_TM_DKP}  using the analysis of the Lorentz group.
Let us add an internal charge index to the tensors. If the internal
charge metric $\kappa_{ab}$ of the charge space
is symmetric, we recover the generalized charged Avdeev-Chizhov
Lagrangian
\BE
\label{eq:2.15}
\LAG^{AC} = \frac{1}{2(p-1)!}\; \kappa_{ab} (\partial . \ZB^a)(\partial . Z^b)
\\
= \frac{1}{ \pp !}\; \kappa_{ab} (-T^a \Box T^b - \DD (\partial . T^a)(\partial . T^b))
\;,\;\;\kappa_{ab} = \kappa_{ba}\;.
\EE
\blue{where ($\partial . Z$) denotes the divergence of $Z$, i.e. a derivation contracted on the first index of the tensor $Z$.}
But we can also construct an antisymmetric Lagrangian, \cite{Thierry_Mieg_Jarvis_2021},
using a \blue{real} antisymmetric metric $\eta_{ij}$ in the internal charge space
\BE
\label{eq:2.16}
\LAG^{CP} = \frac{-i}{(p-1)!}\; \eta_{ij} (\partial . \ZB^i)(\partial . Z^j)
\\
= \;\frac{1}{(p-1)!}\;\eta_{ij} (\partial . {}^*T^i)(\partial . T^j) 
\;,\;\;\eta_{ij} = - \eta_{ji}\;,
\EE
This Lagrangian is real, because complex conjugation switches the sign of the overall factor $i$
but at the same time exchanges $Z$ and $\ZB$ in the definition (\ref{eq:2.12}).
It is pseudo-scalar, because flipping the orientation of space exchanges $(Z,\ZB)$ 
and therefore flips the sign of the Lagrangian.
Yet it naturally implements the Landau $CP$ symmetry because, if the orientation of the internal charge space, the
indices $[ij]$, are flipped at the same time as the orientation of space, the sign of the Lagrangian is maintained.
We call it $CP$ because it implements Landau charge-parity $CP$ symmetry, and also because it is both
$C$:conformally invariant (see below) and $P$:parity aware.

In its $Z$ form, the left hand side of the $\LAG^{AC}$  (\ref{eq:2.15}) is very similar to $\LAG^{CP}$ (\ref{eq:2.16}),
but when we rewrite it as a function of $T$, it is markedly different. In particular the d'Alembertian term, which is necessarily
symmetric, disappears. Thus, to prove the conformal invariance of our Lagrangian $\LAG^{CP}$,
we cannot use the field virials $V^{\Box}$ and $V^{\nabla}$ defined in (\ref{eq:2.8}).
To compute the $V(Z)$ field virial, we start again from
the definition (\ref{eq:2.1}). The advantage is that this new
calculation will cover both the internal charge symmetric $\LAG^{AC}=\LAG^{\chi}_{\kappa}$ 
and the antisymmetric $\LAG^{CP}=\LAG^{\chi}_{\eta}$ cases.
Using $\partial \ZB / \partial Z = 0$, a direct
a consequence of self-duality reminiscent of complex derivatives, we find 
\BE
\label{eq:2.17}
V^{\chi,\nabla}_{\sigma} = \displaystyle{\frac {\partial \LAG^{\chi}}{\partial (\partial_{\alpha}\ZB^i_{\beta\gamma ...})}}\;(\delta g_{\alpha\sigma} - \Sigma_{\alpha\sigma})(\ZB^i_{\beta\gamma ...})
+
\\
+ \displaystyle{\frac {\partial \LAG^{\chi}}{\partial (\partial_{\alpha}Z^j_{\beta\gamma ...})}}\;(\delta g_{\alpha\sigma} - \Sigma_{\alpha\sigma})(Z^j_{\beta\gamma ...})
\\
= \frac{1}{\pp !} (\dd - \DD + \pp)\; \eta_{ij}\;[(\partial_{\alpha}\ZB^{i,\alpha\mu ...})\;Z^j_{\sigma\mu ...} + (\partial_{\alpha}Z^{j,\alpha\mu ...})\;\ZB^i_{\sigma\mu ...}]\;.
\EE
We wish to establish that this field virial is a divergence.
To complete the proof, we take advantage of an interesting identity.
Consider the d'Alembertian $\Box$ of the chiral fields
\BE
\label{eq:2.18}
\LAG^{\chi,\Box} = -\frac{1}{2 \pp!}\;\eta_{ij}\;\ZB^{i,\mu\nu ... }\;\Box\;Z^j_{\mu\nu ...}\;
\EE
In reality, this Lagrangian vanishes because the total contraction of an anti self-dual $\ZB$ with a self-dual tensor $Z$ vanishes.
Now since $Z$ is self-dual, and since $\Box$ does not
alter the tensorial indices, $\Box Z$ is also self-dual and $\LAG^{\chi,\Box} = 0$.
Nevertheless, we can compute the corresponding field virial.
Obviously, it vanishes up to a total derivative, but in a very non trivial way.
\BE
\label{eq:2.19}

V^{Z,\Box}_{\sigma} = -\;\frac{1}{\pp!}\;\eta_{ij} \; (\ZB^{i,\mu\nu ...})\; \partial^{\alpha}\;(\dd g_{\alpha\sigma} - \Sigma_{\alpha\sigma})(Z^j_{\mu\nu ...})
\\
= - \frac{1}{\pp!}\;\eta_{ij} (\ZB^{i,\mu\nu ...})\;(\dd \partial_{\sigma}Z^j_{\mu\nu ...} + \pp (-\partial_{\mu}Z^j_{\sigma\nu ...} +  g_{\mu\sigma}\partial^{\alpha}Z_{\alpha\nu ...}))\;.
\EE
The first term ($\eta \;\ZB\;\dd\;\partial_{\sigma} Z$) vanishes (self-dual to anti-self-dual full contraction), but
the second term does not, so
\BE
\label{eq:2.20}
V^{Z.\Box}_{\sigma} \cong \frac{1}{(\pp -1) !}\;\eta_{ij} (\ZB^{i,\mu\nu ...}\;\partial_{\mu} \;Z^j_{\sigma\nu ...} -  \ZB^{i}_{\sigma\nu ...}\; \partial_{\mu}Z^{j,\mu\nu ...})\;.
\EE
Adding the $\ZB$ derivative we get
\BE
\label{eq:2.21}
V^{\chi,\Box}_{\sigma}  = V^{\ZB,\Box}_{\sigma}  + V^{Z,\Box}_{\sigma}  
\\
=
\frac{1}{(\pp - 1)!}\;\eta_{ij}\; (\ZB^{i,\mu...}\;\partial_{\mu} \;Z^j_{\sigma...} -  \ZB^{i}_{\sigma...}\; \partial_{\mu}Z^{j,\mu ...} +
\\
+ Z^{j,\mu...}\;\partial_{\mu} \;\ZB^i_{\sigma...} -  Z^{j}_{\sigma...}\; \partial_{\mu}\ZB^{i,\mu ...})\;.
\EE
Adding the 2 types of field virials with the proper linear combination, we get a divergence
\BE
\label{eq:2.22}
(\dd - \DD +\pp)V^{\chi,\Box}_{\sigma}  + 2\pp \;V^{\chi,\nabla}_{\sigma} =
\\
(\dd - \DD +\pp)\;\frac{1}{(\pp - 1)!}\;\eta_{ij}\;
\partial_{\mu} (\ZB^i_{\sigma ...} Z^{j,\mu ...} + \ZB^{i,\mu ...} Z^j_{\sigma ...}) \;.
\EE
Since we know a priori that the $V^{\chi,\Box}$ field virial vanishes up to a divergence for the simple reason that the Lagrangian $\LAG^{\chi,\Box}$ vanishes identically,
we conclude that $V^{\chi,\nabla}_{\sigma}$ is a divergence, and therefore the Lagrangian $\LAG^{\chi,\nabla}$ is conformally invariant 
for any choice of the internal charge metric.
In the symmetric case $\LAG^{AC} = \LAG^{\chi,\nabla}_{\kappa}$ where $\kappa_{ab}=\kappa_{ba}$, 
we recover the already demonstrated conformal invariance of the Avdeev-Chizhov Lagrangian.
In the antisymmetric case $\LAG^{CP} = \LAG^{\chi,\nabla}_{\eta}$, where $\eta_{ij}=-\eta_{ji}$, 
we prove that the $CP$ model is conformally invariant.

Fulfilling the classification \cite{Jarvis_TM_DKP}, there exists a second pseudo-scalar Lagrangian, that we can call pseudo-trivial ($PT$)
\BE
\label{eq:2.23}
\LAG^{PT}\;d^{\DD}x = - \frac{1}{2}\;\kappa_{ab}\;T^a\; \Box \wedge T^b
\\
= - \frac{1}{2p!}\;\kappa_{ab}\;\epsilon^{\mu\nu ... \rho\sigma ...}\;T^a_{\mu\nu ...} \;\Box\;T^b_{\rho\sigma ...}\;d^{\DD}x\;.
\EE
This Lagrangian treats the $T_{\mu\nu ...}$ component fields as a collection of scalars. Like $T^{\mu\nu ...} \Box T_{\mu\nu ...} $, it does not recognize the spin of the tensor $T$
and for this reason is not conformally invariant. In $\DD=4k$, it is symmetric in $(\mu\nu ... ,\rho\sigma ...)$, so $\kappa_{ab}$ has to be symmetric in $(ab)$, 
skew in $\DD=4k+2$.

To summarize, we have constructed in $\DD$ dimensions 4 different families of Lagrangians.
Two are invariant ($KR$ and $AC$) and two flip sign if we change the orientation of space
($CP$ and $PT$). Two are using a symmetric 
metric $\kappa_{\{ab\}}$ in the internal charge space ($KR$, $AC$) and one an antisymmetric metric $\eta_{[ij]}$ ($CP$),
$PT$ alternates. 
Two are conformally invariant ($AC$ and $CP$) in any even dimension
and one ($KR$) if and only if the rank $\pp$ of the tensor matches its canonical dimension $\dd$. 
These four cases saturate a Lorentz group-theoretical classification in $\DD=4$ dimensions (see \cite{Jarvis_TM_DKP}, appendix A) which
  proved that there are no other possibilities.
 The corresponding character analysis using representations of the full Lorentz group in arbitrary even dimension D = 2p shows that there remain four one-dimensional candidates, namely two 
$O(2\pp\!-\!1,1)$ scalars in symmetric coupling, plus one $O(2\pp-1,1)$ pseudoscalar in each of symmetric and antisymmetric coupling.
 So the classification is complete in all even dimensions.
 
\section{Physical content of the anti-symmetric tensor theories}
\label{sec:DF}

  A rank  $\pp$  anti-symmetric tensor  in $\DD$ dimensions has $\DD ! \;/\;\pp!\;(\DD - \pp)!$ components.
In the Kalb-Ramond case (\ref{eq:1.2}), because of gauge invariance (\ref{eq:1.3}),
these tensors are massless and their polarization states
are transverse to the light-cone.
For example in $\DD=4$ dimensions, the free $KR$ Lagrangian describes a scalar field,
and \cite{FreedmanTownsend1981} have shown that the non Abelian 
generalization of this $\DD=4$ model 
is equivalent to the non linear $\sigma$-model.
More generally, in $\DD$ dimensions, the number of degrees of freedom carried by a $\pp$-forms 
is equal to the number of components of the same $\pp$-form in the transverse $(\DD-2)$ dimensions.
As verified by \cite{PVN1980}, the $KR$ gauge fields can indeed be quantized using
a triangular pyramid of ghosts: 2 ($\pp-1$)-ghost at level 1, 3 $(\pp-2)$-ghosts at level 2, 4 at level 3, ...
down to $\pp$ scalar ghosts at level $\pp$ alternatively commuting and anti-commuting
according to the ghost combinatorial identity then just discovered in \cite {TM90},
\BE
\label{eq:4.1}
\sum_{i=0}^{\pp} \;(-1)^i (i+1) (^{\pp-i}_{\DD}) = (^{\pp}_{\DD - 2}) \;.
\EE
which coincides with the development of an anti-symmetric Young tableau of the BRS extended
orthosymplectic superalgebra $OSp(D/2)$ of \cite{delbourgo1983exotic} relative to
its maximal even subalgebra $SO(D) \oplus Sp(2)$. 

The situation is more complicated in the Avdeev-Chizhov case (\ref{eq:2.15}, \ref{eq:2.16}).
The tensors $Z$ and $\ZB$ are self dual and anti-self-dual (\ref{eq:2.12})
and each have $\DD ! \; / \;2((\DD/2)!)^2$ components, i.e. 3 components in 4 dimensions.
The free Lagrangian is invertible and the propagators are well defined.
This is consistent with the absence of gauge invariance, since there is no need and no
possibility of adding a gauge fixing term like in the Maxwell or Kalb-Ramond models.
Nevertheless, in 4 dimensions, a real Avdeev-Chizhov tensor carries only one degree of
freedom (\cite{AvdeevChizhov1994B,LemesRenanSorella1995A,LemesRenanSorella1995B,Wetterich2008,Chizhov2011})

Although there is no gauge invariance, \cite{AvdeevChizhov1994B} proposed in 4 dimensions a possible
set of Faddeev-Popov ghosts and BRS equations
used by \cite{KirilovaChizhov2017} in a way reminiscent of \cite{PVN1980}.
An alternative could be to consider again the BRS extended superalgebra
and introduce the infinite tower of ghosts associated
to the $OSp(4/2)$ self-dual supertensor
consisting $\forall n \in \mathbb{N}$ of a self-dual 2-tensor at level $4n$, 
an anti-self-dual 2-tensor at level $4n+2$ and a ghost vector at level $2n+1$.
More generally the superdimension of an $OSp(D/2)$ self-dual supertensor
is equal to the dimension of the self-dual tensor of the space $SO(D-2)$ transverse to
the light cone (\cite {JeugtTmStoilova2017}). This is most likely the number of physical components
of the self dual model in dimension $D$, although an explicit Hamiltonian proof is lacking in $\DD > 4$.

The phenomenology of self-dual tensors in $\DD=4$ dimensions is well detailed in \cite{Chizhov2011}.
Their renormalization is analyzed in 
\cite{LemesRenanSorella1995A,LemesRenanSorella1995B} and \cite{Wetterich2008}.
As shown in the interesting paper of \cite{Shapiro2024},
the curved-space extension of the Avdeev-Chizhov Lagrangian
is locally scale invariant (Weyl conformal invariant).
See also the geometric constructions of \cite{Hughes2024}.

The $CP$ model might be a promising  candidate to construct an extension of the standard model 
because in 4 dimensions, \blue{as previously mentioned}, self-dual 2-tensors naturally couple to chiral Fermions.
Furthermore, as shown in \cite{Thierry_Mieg_Jarvis_2021},
the counter-term to the propagator of the $T$ tensor induced by the Fermion loops is 
constrained by the Bouchiat-Iliopoulos-Meyer
(BIM) mechanism to induce an antisymmetric metric $\eta_{ij}$ in the internal space,
thus favoring the (CP) model over the (AC) model.
The Fermion loops also induce a triple
scalar-vector-tensor vertex $\Phi\;F^{\mu\nu}T_{\mu\nu}$. As this term depends on the derivatives
of the vector field $F_{\mu\nu}=\partial_{[\mu}A_{\nu]}$, it could contribute to the field virial
and break conformal invariance. However
\BE
\label{eq:2.24}
\LAG^{\Phi A B} =\frac{1}{2} \Phi  F^{\mu\nu} T_{\mu\nu} \;,\;\;
\\
V_{\sigma}^{CP} = \displaystyle{\frac {\partial \LAG^{\Phi A T}}{\partial(\partial_{\alpha} A_{\mu})}} \;(\delta\;g_{\alpha\sigma}\;-\Sigma_{\alpha\sigma})\;(A_{\mu}) = (\dd-1) \Phi T_{\sigma\mu}A^{\mu}\;.
\EE
Hence the scalar-vector-tensor vertex $\Phi\;F^{\mu\nu}T_{\mu\nu}$ is conformally invariant in $\dd=1 \Leftrightarrow \DD=4$ dimensions.

\section{Conclusion}
\label{sec:Conclusions}

Conformal invariance is an important symmetry  (see  for example the
physics report of \cite{Nakayama2015}), and we provide for the first time a systematic analysis
of the conformal invariance of antisymmetric-tensors Lagrangian field theories in any even dimension.
First, we have analyzed the Kalb-Ramond gauge tensor Lagrangian and using the conformal
field virial $V_{\sigma}$ (\ref{eq:2.1}) constructed by \cite{Jackiw_2011},
we have shown (\ref{eq:2.11}) that the theory is
conformally invariant if and only if the rank $\pp$ of the gauge tensor is equal to
its canonical dimension $\dd = (\DD - 2)/2$. This case covers the $\pp=0$ scalars of the 2-dimensional string theory,
the $\pp=1$ vector of the 4-dimensional Maxwell-Yang-Mills theory, the $\pp=2$ Kalb-Ramond 2-tensor theory in 6 dimensions
with cubic interactions,
and also 3-tensors in 8 dimensions, 4-tensors in 10 dimensions and so on. This direct proof is new,
compare with \cite {Kuzenko2021,Lee2021}. However, the
Kalb-Ramond 2-form gauge fields are not conformally
invariant in 4 dimensions, and the Maxwell theory is conformally invariant only in 4 dimensions, \cite{Bateman1910}.

We then analyzed the \cite{AvdeevChizhov1994A} matter tensor Lagrangian (\ref{eq:2.7}),
generalized it to higher dimensions
and proved for the first time that the self-dual matter tensor fields are conformally invariant in any even dimension (\ref{eq:2.14}).

Furthermore, by casting the theory into a chiral form (\ref{eq:2.15}),
we constructed in any even dimension a novel family of Lagrangians  (\ref{eq:2.16})
describing charged self-dual tensors which we 
call $\LAG^{CP}$ because they implement the Landau
charge-parity $CP$ invariance, and because we proved that
these Lagrangians are $C$:conformally invariant (\ref{eq:2.22}) and $P$:parity aware.
We first discovered these new tensors in our analysis of the superalgebraic $SU(2/1)$ structure
of quarks and leptons. By analyzing the novel scalar-vector-tensor triangle interaction induced
in that theory by the Fermion loop counter-terms, we prove (\ref{eq:2.24})
that the complete $SU(2/1)$ super-chiral super-connection Lagrangian we
constructed in \cite{Thierry_Mieg_Jarvis_2021} is
conformally invariant, reinforcing the possibility that it could play a role in a superalgebraic extension
of the standard model.

We hope that this focused study on conformal invariance, based on the elegant construction of Jackiw and Pi, will raise interest in self-dual tensors.

\section*{Acknowledgments}
We are grateful to Sergei Kuzenko, Ivan Todorov, and especially Mikhail Chizhov, for comments.
We also thank the referees for their constructive suggestions.
This research was supported in part by the Intramural Research Program of the National Library of Medicine, National Institute of Health.


\appendix

\section{The gauged fixed Maxwell model is conformally invariant in $\DD \ge 4$}
\label{ap:N6}

It is well known that the Maxwell theory is conformally invariant in $\DD=4$ dimensions, \cite{Bateman1910},
or in $\DD=3$ by duality, but not in $\DD>4$
(see for example \cite{Jackiw_2011, Showk2011}).
However, in the same paper El-Showk noticed that the
gauged fixed Maxwell model is conformally invariant
in $\DD > 4$ for a particular gauge choice. That proof can be greatly simplified
using the Jackiw and Pi theorem as follows. Consider the gauged fixed Maxwell Lagrangian
\BE
\LAG^{MGF} = \LAG^M + \LAG^{GF}\;,\;\;\LAG^M = -\frac{1}{4} (F_{\mu\nu})^2\;,\;\;\LAG^{GF}=  -\frac{1}{2\xi}\;(\partial^{\mu}A_{\mu})^2\;.
\EE
We dropped the Faddeev-Popov ghosts, since scalars are conformally invariant.
The Jackiw current associated to the classical Maxwell term has already been computed (\ref{eq:2.6})
\BE
V^M_{\sigma}(A) = \displaystyle{\frac {\partial \LAG^M}{\partial(\partial_{\alpha} A_{\mu})}} \;(\dd\;g_{\alpha\sigma}\;-\Sigma_{\alpha\sigma})\;(A_{\mu}) =
-(\dd - 1) A^{\alpha} (\partial_{\sigma} A_{\alpha}  - \partial_{\alpha} A_{\sigma})\;. 
\EE
The first term is by itself a divergence $A^{\alpha} \partial_{\sigma} A_{\alpha}  = 1/2 \;\partial^{\beta}(\delta_{\beta\sigma}\; A^{\alpha}A_{\alpha})$,
but the second term, which can be rearranged as  $(\dd-1)(\partial^{\alpha}A_{\alpha})A_{\sigma}$ is not.
However, we must also consider the Jackiw current $V^{GF}$ associated to the gauge fixing term
\BE
V^{GF}_{\sigma} =  \displaystyle{\frac {\partial \LAG^{GF}}{\partial(\partial_{\alpha} A_{\mu})}} \;(\dd\;g_{\alpha\sigma}\;-\Sigma_{\alpha\sigma})\;(A_{\mu})
 =-\frac{1}{\xi} (\dd + 1 - \DD) (\partial^{\mu}A_{\mu})\;A_{\sigma}
 \EE
 Therefore if $\xi = (\DD - \dd - 1)/(\dd - 1) = \DD / (\DD - 4)$, the unwanted second term is canceled,
 the total Jackiw current $J^M + J^{GF}$ is a divergence,
 and the model is conformally invariant. We have recovered the \blue{result} of Showk, but it a much simpler way.
 Notice however that conformal invariance only applies to the free theory.
 The canonical dimension of the vector field $A$ is $\dd=(\DD-2)/2$,
 so $d=1$ only in dimension $D=4$.
 Therefore the coupling constant $\gcpl$ in the covariant derivative $D_{\mu} = \partial_{\mu} + \gcpl \; A_{\mu}$ is dimensionless
 only in $\DD=4$, hence the gauge fixed coupled QED or Yang-Mills models are not conformally invariant in $\DD > 4$.

\section{The quartic D=6 vector model is not conformally invariant}
\label{ap:MQ}

  In section C.3 of \cite{Lee2021}, it is claimed that in 6 dimensions the quartic Lagrangian 
\BE
\LAG^{Q} = -\frac{1}{4} \; F^{\mu\nu} \Box F_{\mu\nu}\;,\;\;\; F_{\mu\nu} = \partial_{\mu}A_{\nu} - \partial_{\nu}A_{\mu}\;,
\EE
is conformally invariant. The advantage of this quartic Lagrangian is that in $\DD=6$ the $A_{\mu}$ vector field
has canonical dimension 1, allowing covariant derivatives with a dimensionless coupling constant. However, using Jackiw and Pi,
we can show that this Lagrangian is not conformally invariant
because the $\Box$ d'Alembertian does not take into account the tensorial nature of $F_{\mu\nu}$
and treats it as a collection of 6 scalars.
Indeed, when we compute the derivative $\partial / \partial(\partial_{\alpha} A_{\mu})$, the $\Box$ term is not inert
but contributes to the current and we find
\BE
V^{Q}_{\sigma} = -(\dd - 1) (\partial_{\sigma}\partial_{\mu} (\partial . A))\;A_{\mu} + \dd \; (\partial_{\sigma}\;\Box A_{\mu})\;A_{\mu} - (\partial_{\mu}\;\Box A_{\sigma})\;A_{\mu}
\EE
which is not a divergence. Hence the quartic model $\LAG^{Q}$ is not conformally invariant.

\pagebreak
\bibliographystyle{elsarticle-harv}

\bibliography{Conformal_v2_blue.bib}






\end{document}